\documentclass[reprint,amsmath,amssymb,superscriptaddress,pra]{revtex4-1}

\usepackage[charter,cal=cmcal,sfscaled=false]{mathdesign}

\usepackage{dcolumn}
\usepackage{bm}
\usepackage{dsfont}
\usepackage{graphicx,epsfig}
\usepackage{amsmath}
\usepackage{braket}
\usepackage{physics}
\usepackage{caption}
\usepackage{float}
\usepackage[shortlabels]{enumitem}

\usepackage{blindtext}
\usepackage{color}

\usepackage[colorlinks]{hyperref}
\hypersetup{
	colorlinks = true,
	urlcolor = {blue},
	citecolor = {blue},
	linkcolor= {blue}
}

\newenvironment{proof}{%
  \par\noindent\textit{Proof.}%
}{\hfill$\blacksquare$\par\vspace{0.3cm}}

\renewcommand{\eqref}[1]{Eq.~(\ref{#1})}
\newcommand{\figref}[1]{Fig.~\ref{#1}}

\newcommand{\secref}[1]{Sec.~\ref{#1}}

\newtheorem{result}{Result}
\newtheorem{corollary}{Corollary}

\begin{document}

\title{Entanglement Structure Certification Based on Energy-Restricted State Discrimination}

\author{Carles Roch i Carceller}\email{carles.roch\_i\_carceller@fysik.lu.se}\address{Physics Department and NanoLund, Lund University, Box 118, 22100 Lund, Sweden}

\begin{abstract}
The certification of entanglement in multipartite scenarios is crucial for the advancement of quantum technologies, particularly for the realization of large-scale quantum networks. Here, we introduce a method to certify the structure of the entanglement in ensembles of quantum states with limited energy based on a state discrimination game played by multiple distant and uncharacterized parties. The optimal success probability of this game forms a strict hierarchy, determined by the number of bipartitions and the size of the entangled subsets in each state of the underlying ensemble. The game can be optimally won using a single, fixed measurement setting shared by all parties, regardless of the specific entanglement structure. We further demonstrate that both the performance and noise robustness of our method improve in the multipartite regime, scaling exponentially with the number of parties. Consequently, our approach enables the exclusion of entire structural classes, thereby certifying the structure of multipartite entanglement.
\end{abstract}

\maketitle

\section{Introduction} 

The rapid progress of quantum technologies is steadily transforming the vision of a global quantum internet into a tangible goal. Quantum networks lie at the heart of this endeavour, promising revolutionary capabilities such as unconditionally secure communication~\cite{Bennett1992,Ekert1991}, distributed quantum computation~\cite{Cirac1999,Kimble2008}, and enhanced simulations of complex quantum systems~\cite{Georgescu2014}. Realizing these possibilities, however, requires not only the ability to distribute quantum states over long distances, but also reliable methods to verify that the essential non-classical resources are indeed present. Among these resources, entanglement occupies a central role. It underpins the security of quantum key distribution~\cite{Gisin2002,Scarani2009}, enhances quantum metrology~\cite{Giovannetti2006,Dooley2023,Huang2024,Kuriyattil2025} and many times improves distributed quantum communication complexity tasks~\cite{Cleve1999,Buhrman2001,Brassard2003,Brukner2004}. The ability to certify the presence of entanglement is therefore indispensable, both from a fundamental perspective and for practical deployment of quantum technologies.  

In multipartite settings, certification becomes even more demanding. It is not sufficient to establish the presence of entanglement across some cut of the network; rather, one must assess the full structure of correlations to determine the depth and extent of entanglement shared among all parties \cite{Sorensen2001,Lucke2014,Navascues2020,Frerot2023,Xu2025,Qi2025}. At the extreme, genuine multipartite entanglement (GME)---where entanglement extends across all possible bi-partitions---represents the strongest form of entanglement relevant for quantum networks, and its certification lies at the cornerstone of entanglement detection \cite{Seevinck2001,Guhne2009,Guhne2010,Huber2010,Jungnitsch2011,Huber2011}. It is, however, also crucial to detect intermediate levels of entanglement. The underlying structure of a multipartite entangled state is typically quantified through the notions of separability and producibility \cite{friis2019}. Namely, a multipartite quantum state is said to be $k$-separable if it can be written as a convex mixture of tensor products of $k$ disjoints subsets \cite{Horodecki2009}. On the other hand, a quantum state is said to be $l$-producible if all disjoints states contain at most $l$ particles \cite{Guhne2005}. Together, both separability and producibility typically classify distinct structural classes of multipartite entanglement, and its certification has been central with various approaches. One the one hand, device-independent witnesses \cite{Bancal2011,Moroder2013,Liang2015,Aloy2019} offer a solution to tomography-based methods, which become impractical in the many-body regime. These, however, are based on Bell-like inequalities, and rely on the experimentally demanding task of the detection of non-locality. Alternatively, semi-device-independent approaches emerged, based on restricting the dimension of the underlying quantum systems~\cite{Tavakoli2018,lu2018,Moreno2021}. This approach suffers from the main caveat that the Hilbert space dimension is not a direct measurable quantity. To overcome this issue, here we rely on physically motivated assumptions about the underlying states by considering a restriction on the energy of state preparations \cite{VanHimbeeck2017}. Many realistic implementations---ranging from photonic platforms to atomic systems---are inherently limited in energy, and it is therefore both practical and meaningful to consider certification frameworks that explicitly exploit such restrictions \cite{Reimer2014,Mazeas2016,Martin2017,Wen2022,Kao2024}.  

In this work, we introduce a semi-device-independent task that leverages energy constraints to certify the structure of the entanglement in a network setting. The central idea relies on a state discrimination game, in which energy-restricted ensembles with stronger entanglement structures outperform all of their separable counterparts. Remarkably, since most multipartite state discrimination tasks achieve optimal performance without requiring complex global entangled measurements, our protocol naturally inherits this desirable simplicity~\cite{walgate2000,virmani2001,ji2005,Bakhshinezhad2024,carceller2025}. This performance gap constitutes a direct and operationally meaningful signature of entanglement, allowing one to quantitatively certify the underlying structure of entanglement present in ensembles of quantum states. Importantly, the presented method provides a stronger certification of the entanglement structure which goes beyond current standards, as it can detect entanglement architectures that are indistinguishable by separability and producibility combined. 

This task is illustrated in \figref{fig:simple_qsd_dist}, to which we have coined the term \textit{distributed state discrimination}. In its simplest form, a preparation device receives two classical bits, $x_0$ and $x_1$, and uses a certain amount of energy $\omega$ to encode them into a quantum state. This state is distributed to two distant parties. The goal of each party is to recover their designated input bit: party $0$ aims to infer $x_0$, while party $1$ seeks to infer $x_1$. In \secref{sec:scenario}, we formally introduce the distributed state discrimination task for two parties, and demonstrate that energy-restricted entangled states enable a strictly higher success probability than any ensemble of separable states. Crucially, we show that this entanglement advantage is attained with simple local measurement strategies. In \secref{sec:multipartite} we generalize our findings going beyond the bipartite scenario to a multipartite setting, naturally extending the approach to ensembles of states distributed among $n$ parties. Here we show that the discrimination advantage provided by fully entangled ensembles vs.~fully separable ensembles grows exponentially with the number of parties. Finally, in \secref{sec:certification} we arrive to the practical application of the distributed state discrimination task investigating the performance of ensembles of partially entangled states. The analysis reveals a clear hierarchy in the performance dictated by the structure of the entanglement in the ensemble of states, which directly enables the certification of partial entanglement to the extreme of genuine multipartite entanglement.

\section{Scenario and state discrimination task}
\label{sec:scenario}

\begin{figure}
\centering
\includegraphics[width=0.35\textwidth]{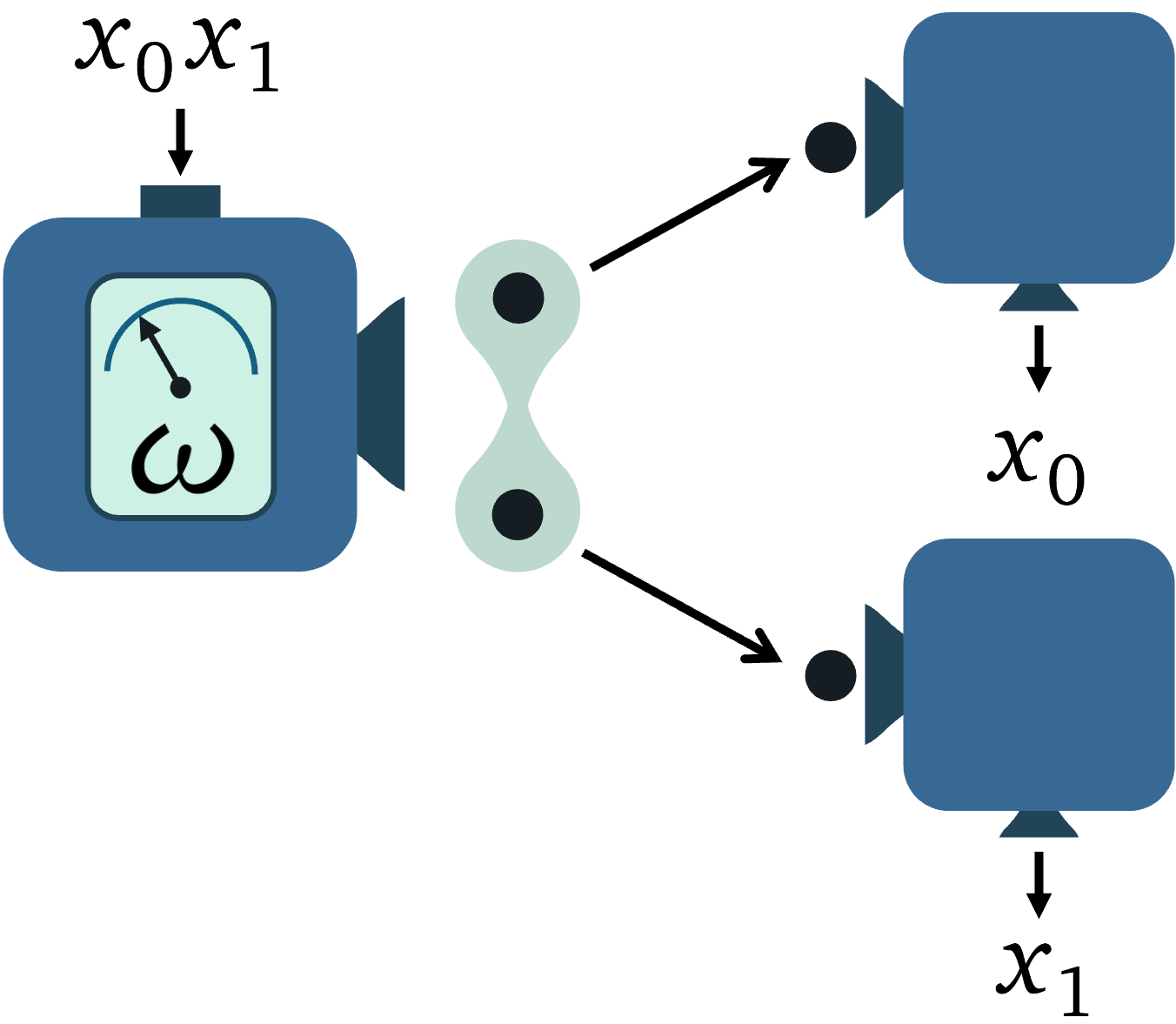}
\caption{\textit{Distributed state discrimination.} A device receives a pair of inputs $\{x_0,x_1\}$ and prepares the bi-partite state $\rho_{x_0 x_1}$. This is then distributed among two receivers $i=\{0,1\}$ which perform a single measurement. The goal of each receiver is to recover the value of $x_i$ in their measurement outcome.}
\label{fig:simple_qsd_dist}
\vspace{-0.5cm}
\end{figure}

We begin by analysing the simplest non-trivial instance of the distributed state discrimination task involving two receivers. The scenario is illustrated in \figref{fig:simple_qsd_dist}. A device encodes two classical bits, $x_0$ and $x_1$, into a bipartite quantum state $\rho_{x_0 x_1}$, which is then distributed among Alice and Bob, who perform local measurements described by POVMs $\{A_a\}$ and $\{B_b\}$, respectively, producing outcomes $a$ and $b$. In general, we allow all devices to be correlated through shared randomness: in each round, a shared variable $\lambda$ is drawn from a distribution $q(\lambda)$, determining the state preparation $\rho_{x_0 x_1}^{(\lambda)}$ and the local measurements $A_a^{(\lambda)}$ and $B_b^{(\lambda)}$. The resulting conditional probabilities are then given by the Born rule,
\begin{align}
p(a,b|x_0,x_1) = \sum_{\lambda}q(\lambda)\Tr\left[\rho_{x_0 x_1}^{(\lambda)} \bigl(A_a^{(\lambda)} \otimes B_b^{(\lambda)}\bigr)\right] \ .
\end{align}
We impose an energy restriction on the states prepared by Alice. As originally proposed in Ref.~\cite{VanHimbeeck2017}, this is formalised by bounding the mean excitation of the prepared quantum states. If we denote the global vacuum by $\ket{\mathbf{0}}:=\ket{0}\otimes\ket{0}$ and consider the Hamiltonian by $H=\openone-\ketbra{\mathbf{0}}{\mathbf{0}}$ with a single non-degenerate ground state and a finite gap, we assume that the energy expectation value of any state in the ensemble is bounded by
\begin{align}\label{eq:E_rest}
\Tr\left[H \rho_{x_0 x_1}\right] \leq \omega \ ,
\end{align}
where $\rho_{x_0 x_1}=\sum_\lambda q(\lambda) \rho_{x_0 x_1}^{(\lambda)}$, for a fixed energy parameter $\omega$. The specific choice of the energy parameter strongly depends on the desired physical implementation. From \eqref{eq:E_rest} it can be inferred that the energy bound can be translated to a bound on a particular component---e.g.~the vacuum or ground state component---of each state in the prepared ensemble. This is a particularly relevant restriction that naturally appears in, e.g., photonic implementations \cite{Weedbrook2012,carceller2025photon} or superconducting circuits \cite{Krantz2019,rasmussen2021}.

The discrimination of pure states under energy-restrictions and without shared randomness has been studied in the single-party setting. There, the maximum probability of correctly identifying each of the $m$ possible state preparations with an energy lower than $\omega$ is upper-bounded by \cite{pauwels2025,carceller2025}
\begin{align} \label{eq:psd}
\mathcal{W}_{m}(\omega) =  \frac{1}{m}\Bigl(\sqrt{\omega(m-1)}+\sqrt{1-\omega}\Bigr)^2 \ .
\end{align}
Here we are interested in a different question. Namely, we investigate how well can Alice and Bob simultaneously recover their individual symbols ($x_0$ for Alice and $x_1$ for Bob) from the set of distributed states $\{\rho_{x_0 x_1}\}$. The figure of merit of the distributed state discrimination task is then the average probability of success,
\begin{align} \label{eq:ps_2parties}
p_s = \frac{1}{4}\sum_{x_0,x_1} \sum_{\lambda}q(\lambda)\Tr\left[\rho_{x_0 x_1}^{(\lambda)} \bigl(A_{x_0}^{(\lambda)}\otimes B_{x_1}^{(\lambda)}\bigr)\right] \ ,
\end{align}
where Alice and Bob succeed if their outcomes match their respective inputs. 
Our first goal now is to determine the maximum achievable value of $p_s$, optimizing over all measurements implementable by Alice and Bob and over all state preparations that obey the energy constraint. However, since shared randomness allows for infinitely many strategies $\lambda$, this optimisation in general can become cumbersome and practically intractable. To simplify the analysis, we proceed showing that shared randomness can be eliminated without loss of generality.  \\

\subsection{Dealing with shared randomness}

The set of reproducible probabilities in energy-restricted semi-device-independent prepare-and-measure frameworks with two inputs and two outputs is known to be closed under shared randomness \cite{VanHimbeeck2017}. This means that any observable correlation attainable with shared randomness and arbitrary energy-restricted quantum states can be reproduced with only pure states without shared randomness. Here, we show that this statement remains true for the set of state discrimination success probabilities with an arbitrary number of inputs and outputs, summarised in the following result.

\begin{result}\label{res:SR}
Consider an ensemble of quantum states $\{\rho_x^{(\lambda)}\}_{x=0}^{m-1}$ such that $\bra{\mathbf{0}}\rho_{x}\ket{\mathbf{0}}\geq 1-\omega$ for $\rho_x=\sum_\lambda q(\lambda) \rho_x^{(\lambda)}$ and an $m$-outcome measurement described through the POVM $\{M_b^{(\lambda)}\}_{b=0}^{m-1}$ sampled according to the distribution $q(\lambda)$. Then, the set of success probabilities $p_s = \frac{1}{m}\sum_{x,\lambda} q(\lambda) \Tr\left[\rho_x^{(\lambda)} M_x^{(\lambda)} \right]$ is closed under shared randomness.
\end{result}

In what follows, we show this result. It will be conceptually important to first consider the case of pure states without shared randomness and describe the whole set of reproducible success probabilities under the energy constraint. Then, we will show that the analogous set of probabilities with shared randomness and arbitrary quantum states is indeed equivalent.

\subsubsection{Pure state preparations without shared randomness}
  
Let us define the set of state discrimination success probabilities attainable with pure state preparations without shared randomness and under energetic restrictions as
\begin{align}
\mathcal{P}_{\omega} = \left\{ p_s \ \big| \  p_s = \frac{1}{m}\sum_{x}\Tr\left[\rho_x M_x \right], \Tr\left[\rho_x H \right]\leq \omega \right\} \ .
\end{align}
We know that any probability $p_s$ that belongs to the set $\mathcal{P}_{\omega}$ is strictly limited by the bound in \eqref{eq:psd} for $0 \leq \omega \leq 1-m^{-1}$ and $p_s\leq 1$ otherwise. Another way of writing the same restrictions is that any attainable success probability $p_s$ must be limited by \cite{pauwels2025}
\begin{align} \label{eq:g_cts}
g_{-}(p_s)\leq \omega \leq g_{+}(p_s) \quad \text{if} \ 0 \leq \omega \leq 1-m^{-1} \ ,
\end{align}
for 
\begin{align}
g_{\pm}(p_s):=\frac{1+(m-2)p_s \pm 2\sqrt{ p_s(1-p_s)(m-1)}}{m}
\end{align}
and $0\leq p_s\leq 1$ otherwise. We can now use the functions $g_{\pm}(p_s)$ to define the set $\mathcal{P}_{\omega}$. To do that though, we need to see that both limits \eqref{eq:psd} and \eqref{eq:g_cts} complement each other. In the energetic range $1-m^{-1}\leq \omega \leq 1$ the implication \ref{eq:psd}$\Leftrightarrow$\ref{eq:g_cts} is trivial. We therefore focus on the range $0\leq\omega\leq 1-m^{-1}$. To see the forward direction \ref{eq:psd}$\Rightarrow$\ref{eq:g_cts}, 
\begin{align}
&g_{-}\left(\frac{1}{m}\left(\sqrt{1-\omega}+\sqrt{\omega (m-1)}\right)^{2} \right) = \omega \\
&\leq g_{+}\left(\frac{1}{m}\left(\sqrt{1-\omega}+\sqrt{\omega (m-1)}\right)^{2} \right)  \ ,
\end{align}
where the second inequality always holds since $g_{-}(p_s)\leq g_{+}(p_s)$, $\forall p_s$. On the other hand, the reverse direction \ref{eq:psd}$\Leftarrow$\ref{eq:g_cts} also holds since,
\begin{align}
p_s &\leq \frac{1}{m}\left(\sqrt{1-g_{-}(p_s)}+\sqrt{g_{-}(p_s) (m-1)}\right)^{2} \\
&\leq \frac{1}{m}\left(\sqrt{1-\omega}+\sqrt{\omega (m-1)}\right)^{2} \\
& \leq \frac{1}{m}\left(\sqrt{1-g_{+}(p_s)}+\sqrt{g_{+}(p_s) (m-1)}\right)^{2}  \ .
\end{align}
Thus, the set $\mathcal{P}_{\omega}$ in the relevant range $0\leq \omega \leq 1-m^{-1}$ is fully described by success probabilities $p_s$ satisfying  $g_{-}(p_s)\leq \omega \leq g_{+}(p_s)$.   \\

\subsubsection{Arbitrary state preparations with shared randomness}
 
Let us now define the set of success probabilities attainable with arbitrary (in the sense that can be pure or mixed) energy-restricted state preparations and shared randomness. This is, we consider realizations $\lambda$ distributed according to $q(\lambda)$ constrained by energies $\omega^\lambda$ that reach of success probabilities $p_s^\lambda$. Then, we define the set of averaged realisations as
\begin{align}
\mathcal{P}'_{\omega} = \left\{ \mathbf{p_s}=\sum_\lambda q(\lambda) p_s^\lambda \ \big| \  p_s^\lambda \in \mathcal{P}_{\omega^\lambda}, \sum_\lambda q(\lambda) \omega^\lambda = \omega \right\} \ .
\end{align}
We now want to show that $\mathcal{P}_{\omega} = \mathcal{P}'_{\omega}$, i.e.~that the set of quantum state discrimination success probabilities is closed under shared randomness. One the one hand, we know that $\mathcal{P}'_{\omega} \supseteq \mathcal{P}_{\omega}$, since any realisation in $\mathcal{P}'_{\omega}$ corresponds to a convex combination of realisations in $\mathcal{P}_{\omega}$. What we need to prove is that $\mathcal{P}'_{\omega} \subseteq \mathcal{P}_{\omega}$, i.e.~that given any mixture $\mathbf{p_s}=\sum_\lambda q(\lambda) p_s^\lambda$ of pure-state success probabilities with averaged energy $\sum_\lambda q(\lambda) \omega^\lambda = \omega$, then the exact same success probability can be obtained through a single pure-state quantum realization with energy $\omega$. This simply amounts to showing that $g_{-}(\mathbf{p_s})\leq \omega \leq g_{+}(\mathbf{p_s})$. And indeed, 
\begin{align}
 g_{+}(\mathbf{p_s}) &= \frac{1+(m-2)\mathbf{p_s} + 2\sqrt{ \mathbf{p_s}(1-\mathbf{p_s})(m-1)}}{m}  \\
 & \geq \sum_\lambda q(\lambda) \frac{1+(m-2)p_s^\lambda + 2\sqrt{ p_s^\lambda(1-p_s^\lambda)(m-1)}}{m} \nonumber \\
 & = \sum_\lambda q(\lambda) g_{+}(p_s^\lambda) \geq \sum_\lambda q(\lambda) \omega^\lambda = \omega \nonumber
\end{align}
\begin{align}
 g_{-}(\mathbf{p_s}) &= \frac{1+(m-2)\mathbf{p_s} - 2\sqrt{ \mathbf{p_s}(1-\mathbf{p_s})(m-1)}}{m}  \\
 & \leq \sum_\lambda q(\lambda) \frac{1+(m-2)p_s^\lambda - 2\sqrt{ p_s^\lambda(1-p_s^\lambda)(m-1)}}{m} \nonumber \\
 & = \sum_\lambda q(\lambda) g_{-}(p_s^\lambda) \leq \sum_\lambda q(\lambda) \omega^\lambda = \omega \nonumber
\end{align}
where we used the Cauchy-Schwartz inequality
\begin{align}
 \sqrt{ \left(\sum_\lambda q(\lambda)p_s^\lambda\right) \left(\sum_\lambda q(\lambda) (1-p_s^\lambda )\right)} \geq \sum_\lambda q(\lambda) \sqrt{p_s^\lambda(1-p_s^\lambda)}  \ .
\end{align}
Therefore, both sets  $\mathcal{P}_{\omega} = \mathcal{P}'_{\omega}$ are equivalent. \\

The distributed state discrimination task we investigate in this work can be viewed as a specialized sub-case of the general state discrimination scenario. Specifically, by mapping the input label $x\rightarrow (x_0,x_1)$ and considering separable measurements $M_b \rightarrow A_a \otimes B_b$, we recover the distributed state discrimination task defined by the success probability in \eqref{eq:ps_2parties}. Consequently, Result~\ref{res:SR} implies that the optimal success probability for the distributed state discrimination task can be determined without considering shared randomness.

\subsection{Separable strategies}

We begin analysing the maximum success probability from \eqref{eq:ps_2parties} achievable when Alice's preparations are restricted to separable states. In this case, the relevant states are of the bi-separable form $\rho_{x_0 x_1}=\sum_\lambda q(\lambda)\psi_{x_0}^\lambda \otimes \phi_{x_1}^\lambda$,
subject to the global energy constraint $\sum_\lambda q(\lambda)\bra{0}\psi_{x_0}^\lambda\ket{0}\bra{0}\phi_{x_1}^\lambda\ket{0}\geq 1-\omega$. Since the set of separable density matrices is convex, and the success probability $p_s$ is a linear functional, the maximum is attained at an extreme point of this set \cite{Krein1940,rockafellar1970,nielsen2000}. Combining this with the fact that the set $\mathcal{P}_{\omega}$ is closed under shared randomness, it suffices to consider pure product preparations of the form $\rho_{x_0 x_1} = \psi_{x_0} \otimes \phi_{x_1}$. With this, the success probability in \eqref{eq:ps_2parties} factorizes, resulting in the following optimisation problem
\begin{align} \label{eq:sep_ps_ini_edit}
p_s^{\rm Sep} = && \underset{\{\psi_{x_0}\},\{\phi_{x_1}\},\{A_a\},\{B_b\}}{\text{maximize}} & \ \frac{1}{4}\sum_{x_0,x_1}\Tr\left[\psi_{x_0} A_{x_0}\right] \Tr\left[\phi_{x_1} B_{x_1}\right] \nonumber \\
	&& \text{such that} \ \ \ \ \ & \ \Tr\left[(\psi_{x_0} \otimes \phi_{x_1})H\right] \leq \omega \ . 
\end{align}
The optimisation is performed over any valid set of quantum states $\{\psi_{x_0}\}$ and $\{\phi_{x_1}\}$---i.e.~normalised $\Tr\left(\psi_{x_0}\right)=\Tr\left(\phi_{x_1}\right)=1$ and positive semidefinite $\psi_{x_0}\succeq 0$, $\phi_{x_1}\succeq 0$, $\forall x_0,x_1$---that satisfy the energy constraint; and POVMs $\{A_a\}$ and $\{B_b\}$ that are complete $\sum_a A_a = \mathds{1}$, $\sum_b B_b = \mathds{1}$ and positive semidefinite $A_a \succeq 0$, $B_b\succeq 0$, $\forall a,b$.

To solve this optimisation problem, let us define the individual success probabilities in Alice and Bobs sides as $p_s^{A}=\frac{1}{2}\sum_{x_0}\Tr\left[\psi_{x_0} A_{x_0}\right]$ and $p_s^{B}=\frac{1}{2}\sum_{x_1}\Tr\left[\phi_{x_1} B_{x_1}\right]$ respectively. Let us further define the individual vacuum components $\bra{0}\phi_{x_0}\ket{0}:=1-\omega_{A}$ and $\bra{0}\psi_{x_1}\ket{0}:=1-\omega_{B}$. According to \eqref{eq:psd}, the maximum achievable success probability for Alice and Bob to correctly discriminate their symbols is limited by $p_s^A \leq \mathcal{W}_{2}(\omega_A)$ and $p_s^B \leq \mathcal{W}_{2}(\omega_B)$. Both individual energy parameters $\omega_A$ and $\omega_B$ are related through the global energy limit constraint in \eqref{eq:sep_ps_ini_edit}, which translates to the vacuum component constraint $(1-\omega_{A})(1-\omega_{B})\geq 1-\omega$. With that, we can re-write the optimisation problem from \eqref{eq:sep_ps_ini_edit} as
\begin{align} \label{eq:sep_ps_ini_edit_2}
p_s^{\rm Sep} = && \underset{\omega_A,\omega_B}{\text{maximize}} & \quad \mathcal{W}_{2}(\omega_A)\mathcal{W}_{2}(\omega_B) \nonumber \\
	&& \text{such that}  & \quad (1-\omega_{A})(1-\omega_{B})\geq 1-\omega \ . 
\end{align}
The function $\mathcal{W}_{n}(\omega)$ is monotonically increasing, i.e.~its maximum is reached when $\omega$ is maximum and thus, the maximum of \eqref{eq:sep_ps_ini_edit_2} is achieved when the inequality constraint is saturated. This optimisation problem can therefore now be solved analytically, substituting $\omega_A$ and nullifying the first derivative of the objective function. This results in 
\begin{align}\label{eq:ps_sep_2parties}
p_s^{\rm Sep} = \frac{1}{4}\Big(1+2\sqrt{(1-\omega)^{1/2}}\sqrt{1-(1-\omega)^{1/2}}\Big)^2 \ ,
\end{align}
with optimal individual energy parameters $\omega_A = \omega_B = 1-\sqrt{1-\omega}$. The maximum in \eqref{eq:ps_sep_2parties} can be simply achieved with the state preparations
\begin{align}\label{eq:2party_states_sep_edit}
\ket{\Psi_{x_0 x_1}} &= (Z_A)^{x_0}\otimes (Z_B)^{x_1} \ket{\Phi^{\rm Sep}_2} \ ,
\end{align}
where
\begin{align}\label{eq:2party_states_sep_edit_2}
\ket{\Phi^{\rm Sep}_2} &= \left(\sqrt{(1-\omega)^{1/2}}\ket{0} + \sqrt{1-(1-\omega)^{1/2}}\ket{1}\right)^{\otimes 2} ,
\end{align}
and the measurements $A_{0}=B_{0}=\ketbra{+}{+}$ and $A_{1}=B_{1}=\ketbra{-}{-}$.

\subsection{Entangled strategies}

After finding the maximum success probability achievable by separable states, we now derive the maximum of \eqref{eq:ps_2parties} when the states are not restricted to be separable. Having established that the set $\mathcal{P}_{\omega}$ is closed under shared randomness, the goal of this task is to find the solution of the following optimisation problem,
\begin{align}
	p^{\rm Ent}_s = && \underset{\{\rho_{x_0 x_1}\},\{A_a\},\{B_b\}}{\text{maximize}} & \quad \frac{1}{4}\sum_{x_0, x_1} \Tr\left[\rho_{x_0 x_1}\left(A_{x_0}\otimes B_{x_1}\right)\right] \nonumber \\
	&& \text{such that} \ \ \ & \quad \Tr\left[\rho_{x_0 x_1}H\right] \leq \omega \ . \label{eq:opt_prob}
\end{align}
Similarly as in \eqref{eq:sep_ps_ini_edit}, the maximisation is performed over all valid quantum state operators that satisfy the energy constraint; and valid POVMs $\{A_a\}$ and $\{B_b\}$. Given that the distributed state discrimination task represents a sub-task of quantum state discrimination, any success probability of the form in \eqref{eq:ps_2parties} will always be lower than $\mathcal{W}_{4}(\omega)$ from \eqref{eq:psd}. Therefore, in order to solve the optimisation problem from \eqref{eq:opt_prob}, we will use an ansatz realisation which can reach the upper-bound from \eqref{eq:psd}. To this end, consider the ensemble of pure state preparations $\rho_{x_0 x_1}=\ketbra*{\Psi_{x_0 x_1}}{\Psi_{x_0 x_1}}$ of the form
\begin{align}\label{eq:2party_states_ent_edit}
\ket{\Psi_{x_0 x_1}} &= (Z_A)^{x_0}\otimes (Z_B)^{x_1} \ket{\Phi^{\rm Ent}_2} \ ,
\end{align}
where $Z_A$ and $Z_B$ respectively denote the Pauli-$Z$ operators in Alice and Bob's systems, and
\begin{align}\label{eq:2party_states_ent_edit_2}
\ket{\Phi^{\rm Ent}_2} &= \sqrt{1-\omega}\ket{00} + \sqrt{\frac{\omega}{3}}\left(\ket{01}+\ket{10}+\ket{11}\right) \ .
\end{align}
The choice of this ensemble of states is motivated by the symmetric structure of the optimisation in \eqref{eq:opt_prob}. Given that the only distinguishability restriction is placed as a bound on the vacuum component of the state preparations, we consider states that are all equally close to the global vacuum $\ket{00}$, and symmetrically spread out around it. Furthermore, we consider that the measurement choices in both Alice and Bob's devices are represented by the POVM elements $A_{0}=B_{0}=\ketbra{+}{+}$ and $A_{1}=B_{1}=\ketbra{-}{-}$.

With this choice of implementation, one can now compute the success probability in the distributed discrimination task, resulting in
\begin{align}\label{eq:ps_ent_2parties}
p_s^{\rm Ent} = \frac{1}{4}\Big(\sqrt{3\omega}+\sqrt{1-\omega}\Big)^2 \ .
\end{align}
This coincides exactly with the upper-bound $\mathcal{W}_{4}(\omega)$ for discriminating four energy-restricted quantum states from \eqref{eq:psd}, thereby proving that the bound is tight.

The states from \eqref{eq:2party_states_ent_edit} that reach the distributed state discrimination success probability from \eqref{eq:ps_ent_2parties} are entangled for any value of $\omega$. One can verify this analysing the spectrum of $\Tr_A(\rho_{x_0x_1})$. The spectrum of the partial transpose is preserved under unitary operations, such as those that define the states in \eqref{eq:2party_states_ent_edit}. It is thus enough to analyse only one state, for instance $\rho_{0 0}$. The spectrum of $\Tr_A(\rho_{0 0})$ contains the eigenvalue $-\frac{1}{3}\sqrt{\omega \tau}$, with $\tau=3-2(\sqrt{3\omega(1-\omega)}-\omega)$, which is negative for any value of $\omega$. Therefore, the states in the ensemble have a negative semidefinite partial-transpose (i.e.~are not PPT) and thus, are entangled \cite{peres1996,Horodecki1996}. \\

\begin{figure*}
\centering
\includegraphics[width=\textwidth]{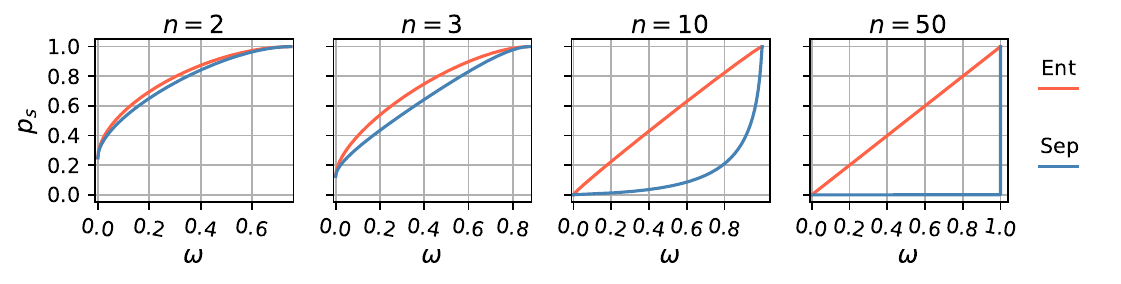}
\vspace{-0.5cm}
\caption{\textit{Entanglement advantage in distributed state discrimination.} Maximum success probability $p_s$ for all $n$ parties to simultaneously discriminate their corresponding bits from ensembles of states with energy $\omega$. Implementations with entangled state preparations (Ent) are able to reach success probabilities higher than fully separable states (Sep).}
\label{fig:ps}
\end{figure*}

Strategies based on entangled state preparations strictly outperform all implementations employing ensembles of separable states, i.e.~$p_s^{\rm Ent} \geq p_s^{\rm Sep}$, for $0<\omega < \frac{3}{4}$, and are only equivalent on the extremes $\omega=0$ and $\omega \geq \frac{3}{4}$. To see this, one might simply check the quantities inside the squares from \eqref{eq:ps_ent_2parties} and \eqref{eq:ps_sep_2parties}, or equivalently, making the change $\sqrt{1-\omega}=t$ and evaluate whether if the function $F(t):=\sqrt{3(1-t^2)}+t-1-2\sqrt{t(1-t)}$ is positive. Indeed, one finds that $F(t)\geq 0$ in the domain $t\in(\frac{1}{2},1)$, saturating the equality on the extremes. In \figref{fig:ps} (leftmost plot) we show the maximum success probabilities achievable with entangled and separable strategies distributed among $n=2$ distant parties. Although subtle, there is a gap between both bounds---$p_s^{\rm Sep} < p_s^{\rm Ent}$ in the relevant energetic range $0<\omega<\frac{3}{4}$---which serves as a signature of entanglement.

\section{Multipartite distributed state discrimination}
\label{sec:multipartite}

So far, we have considered the task of distributed state discrimination among two parties. We now generalise our analysis to an arbitrary number of measuring parties $n$, each aiming to discriminate its corresponding binary message $x_i\in\{0,1\}$. Consider a multipartite ensemble of $m=2^n$ state preparations $\{\rho_{\vec{x}}\}$, each satisfying the energetic restriction $\Tr\left[H\rho_{\vec{x}}\right]\leq \omega$. The overall success probability in the distributed state discrimination game is
\begin{align}\label{eq:ps_multi}
p_s=\frac{1}{2^n} \sum_{\vec{x}} \Tr\left[\rho_{\vec{x}} \left(\bigotimes_{i=0}^{n-1} M_{x_i}^i\right)\right] \ ,
\end{align}
where $\vec{x}=(x_0,x_1,\ldots,x_{n-1})$ labels the global message, and $M_{b_i}^i$ denotes the POVM element corresponding to outcome $b_i$ in the laboratory of the $i^{\text{th}}$ party.

As in the bipartite case, we begin optimising the success probability from \eqref{eq:ps_multi} by considering fully separable state preparations restricted by their energy content. This leads us to the following result.

\begin{result} \label{result2}
Consider a device that prepares $2^n$ equiprobable quantum states  $\rho_{\vec{x}}=\bigotimes_{i=0}^{n-1} \sigma_{x_i}^i$ with $x_i\in\{0,1\}$, such that $\Tr\left[H \rho_{\vec{x}}\right]\leq \omega$, and distributes them among $n$ distant parties. The maximum probability that each party successfully discriminates their corresponding symbol $x_i$ is
\begin{align} \label{eq:ps_sep}
p_s^{\rm Sep}(\omega,n) = \left(\frac{1}{2}+\sqrt{(1-\omega)^{1/n}}\sqrt{1-(1-\omega)^{1/n}}\right)^n .
\end{align}
\end{result}
\begin{proof}
Considering state preparations which are separable across all possible bi-partitions implies that the optimal success probability reduces to the product of the individual success probabilities, 
\begin{align}\label{eq:ps_multi_sep}
p_s=\frac{1}{2^n} \sum_{\vec{x}} \Tr\left[\left(\overset{n-1}{\underset{i=0}{\bigotimes}} \sigma_{x_i}^i\right) \left(\bigotimes_{i=0}^{n-1} M_{x_i}^i\right)\right] = \prod_{i=0}^{n-1}p_s^i \ ,
\end{align}
where we denoted $p_s^i=\frac{1}{2}\sum_{x_i}\Tr\left[\sigma_{x_i}^i M_{x_i}^i\right]$ the individual success probability of the $i^{th}$ party. Let us call $1-\tilde{\omega}_i:=\bra{0}\sigma_{x_i}^i\ket{0}$ the vacuum component of each individual share of the original state. We know that each individual success probability $p_s^i$ with a fixed energy parameter $\tilde{\omega}_i$ cannot be larger than $\mathcal{W}_{2}(\tilde{\omega}_i)$ form \eqref{eq:psd}. To find the maximum achievable success probability with fully separable state preparations, we thus need to solve the following optimisation problem,
\begin{align} \label{eq:full_sep_ps}
p_s^{\rm Sep}(\omega,n) = && \underset{\{\tilde{\omega}_i\}}{\text{maximize}} & \quad \prod_{i=0}^{n-1}\mathcal{W}_{2}(\tilde{\omega}_i) \nonumber \\
	&& \text{such that}  & \quad \prod_{i=0}^{n-1}(1-\tilde{\omega}_{i})\geq 1-\omega \ . 
\end{align}
This optimisation problem can be trivially solved if one realises that, due to the symmetry of the problem, at the optimal point all individual energy parameters $\tilde{\omega}_i$ are equivalent. Indeed, let us call $P=\prod_{i=0}^{n-1}\mathcal{W}_{2}(\tilde{\omega}_i)$ and $K=\prod_{i=0}^{n-1}(1-\tilde{\omega}_{i})$, and set up the Lagrangian $\mathcal{L} = P - \lambda \left( K - (1-\omega) \right)$. To find stationary points, we take the partial derivative with respect to $\tilde{\omega}_i$ and set it to zero. With the appropriate rearranging of terms, one finds
\begin{align} \label{eq:lagrangian_eqs}
\frac{\partial \mathcal{L}}{\partial \tilde{\omega}_i} = 0 \quad \Longrightarrow \quad (1-\tilde{\omega}_i)\frac{\mathcal{W}_{2}'(\tilde{\omega}_i)}{\mathcal{W}_{2}(\tilde{\omega}_i)} = -\lambda \frac{K}{P} \ \forall i ,
\end{align}
where $\mathcal{W}_{2}'(\tilde{\omega}_i)$ denotes the full derivative. On the left hand-side of the resulting identity we have a monotonic function of $\tilde{\omega}_i$ equating a constant term, implying that all $\tilde{\omega}_i$ are equivalent at the optimal point. With that, we substitute $\tilde{\omega}_i=1-(1-\omega)^{1/n}$ in the objective function from \eqref{eq:full_sep_ps} and obtain the desired result in \eqref{eq:ps_sep}.
\end{proof}

The separable implementation that reaches maximum success probability in \eqref{eq:ps_sep} is simply
\begin{align}\label{eq:n_party_state_sep}
\ket{\Psi_{\vec{x}}} &= \bigotimes_{i=0}^{n-1} Z^{x_i}_i \ket{\Phi^{\rm Sep}_n} \ ,
\end{align}
where
\begin{align}
\ket{\Phi^{\rm Sep}_n} &= \left(\sqrt{(1-\omega)^{1/2}}\ket{0} + \sqrt{1-(1-\omega)^{1/2}}\ket{1}\right)^{\otimes n} .
\end{align} 
We now continue by finding the maximum success probability from \eqref{eq:ps_multi} when the states are not constrained to be separable. Already in the $n=2$ parties case, we saw that ensembles of entangled states are more distinguishable than separable states. We thus now investigate the general case of ensembles of $m=2^n$ states distributed among $n$ distant parties, which leads us to the following result.

\begin{result}
\label{result1}
Consider a device that prepares $2^n$ equiprobable quantum states $\rho_{\vec{x}}$ with $x_i\in\{0,1\}$, such that $\Tr\left[H \rho_{\vec{x}}\right]\leq \omega$, and distributes them among $n$ distant parties. The maximum probability that each party successfully discriminates their corresponding symbol $x_i$ is
\begin{align} \label{eq:ps_ent}
p_s^{\rm Ent}(\omega,n) = \frac{1}{2^n}\left(\sqrt{(2^n-1)\omega}+\sqrt{1-\omega}\right)^2 \ .
\end{align}
\end{result}
\begin{proof}
Given that the distributed state discrimination task represents a sub-task of quantum state discrimination, the success probability in \eqref{eq:ps_multi} cannot be larger than $\mathcal{W}_{2^n}(\omega)$ from \eqref{eq:psd}. Therefore, to find the maximum of \eqref{eq:ps_multi} it will suffice to show that it can reach $\mathcal{W}_{2^n}(\omega)$. To this end, we consider $2^n$-dimensional pure state preparations with a vacuum state component $\sqrt{1-\omega}$ that are symmetrically distributed among all parties. This leads to the states $\rho_{\vec{x}}=\ketbra*{\Psi_{\vec{x}}}{\Psi_{\vec{x}}}$, with
\begin{align}\label{eq:n_party_state}
\ket{\Psi_{\vec{x}}} &= \bigotimes_{i=0}^{n-1} Z_i^{x_i} \ket{\Phi^{\rm Ent}_n} \ ,
\end{align}
where $Z_i$ is the Pauli-$Z$ operator acting on the $i^{\rm th}$ party, and
\begin{align}\label{eq:n_ent_state}
\ket{\Phi^{\rm Ent}_n} = \sqrt{1-\omega}\ket{\mathbf{0}} + \sqrt{\frac{\omega}{2^{n}-1}}\sum_{i,j}\ket{\vec{\nu}_{i}^{j}} \ ,
\end{align}
where $\vec{\nu}_{i}^{j}$ denotes a vector full of $i>0$ ones and $n-i$ zeros, and $j$ indexes the possible different permutations without repetitions. Furthermore, we consider that all parties perform the same projective measurements found to be optimal in the bipartite case, i.e.~$M_{0}^{i}=\ketbra{+}{+}$ and $M_{1}^{i}=\ketbra{-}{-}$, $\forall i$. Putting all together leads to the success probability
\begin{align}
&p_s^{\rm Ent}(\omega,n) = \\
&\frac{1}{2^n}\left(\sqrt{1-\omega} + \frac{\omega}{2^n-1}\sum_{i=1}^{n-1}\begin{pmatrix}
n \\
i
\end{pmatrix} + \sqrt{\omega - \sum_{i=1}^{n-1}\begin{pmatrix}
n \\
i
\end{pmatrix} \frac{\omega}{2^n-1}}\right)^2 \nonumber .
\end{align}
The sums over the combinatory elements straightforwardly yield $\sum_{i=1}^{n-1}\begin{pmatrix}
n \\
i
\end{pmatrix} = 2^n - 2$. The final success probability results in \eqref{eq:ps_ent}, which is exactly the maximum state discrimination success probability $\mathcal{W}_{2^n}(\omega)$ from \eqref{eq:psd} with $m=2^n$ pure quantum states with energy upper bounded by $\omega$. 
\end{proof}

Result \ref{result1} shows two key features of this task. First, entangled states lead to better performance in the distributed state discrimination task for an arbitrary number of measuring parties. One can easily verify that the state in \eqref{eq:n_ent_state}, and also the unitarily equivalent states from \eqref{eq:n_party_state}, are entangled for $0 < \omega < 1-2^{-n}$. Namely, by checking the spectrum of the partial transpose over any possible bi-partition of $k$ and $n-k$ qubits is non-positive for any value of $\omega$, as it contains the eigenvalue $-\sqrt{\frac{\omega(2^k-1)(2^{n-k}-1)}{2^n-1}}\abs{\sqrt{1-\omega}-\sqrt{\frac{\omega}{2^n-1}}}$. Second, that the optimal measurement to discriminate a set of $m$ quantum states is separable across any partition that measures the original state, i.e.~$\bigotimes_{i=0}^{n-1} M_{b_i}^i$. That entangled and separable measurement strategies are equally powerful in quantum state discrimination is something that was already observed, e.g.~see Refs.\cite{walgate2000,virmani2001,ji2005,carceller2025}. Moreover, independently of whether the ensembles are entangled or not, the distributed state discrimination game is optimally won when all parties dispose of the exact same single measurement setting---for what in our construction it corresponds to the projections $M_{\pm}^{i}=\ket{\pm}\bra{\pm}$, $\forall i$. This remark will extend to even intermediate cases of multipartite entanglement, showcasing the feasible scaling of the method with the number of parties.

In \figref{fig:ps} we show the success probability bounds for both entanglement-based and fully separable strategies in the cases of $n=2,3,10$ and $50$ parties. The plots reveal a clear gap in the energy range $0<\omega<\omega^\ast$, with $\omega^\ast=1-2^{-n}$. This gap becomes particularly pronounced in the multipartite regime: as $n$ grows, the separable bound rapidly approaches zero, while the entanglement-based bound converges towards the linear behaviour $\omega$. In this context, we address the noise tolerance of entanglement advantage in the asymptotic limit. Specifically, for quantum states subject to standard white noise of the form $\rho_{\vec{x}}^\nu = \nu \rho_{\vec{x}} + \frac{1-\nu}{2^n}\mathds{1}$, we compute the visibility $\nu$ at which the success probability for $\rho_{\vec{x}}^{\nu}$ meets $p_s^{\rm Sep}(\omega,n)$, leading to
\begin{align}
\nu_{\rm crit}^{\rm Ent} = \frac{p_s^{\rm Sep}(\omega,n) - 2^{-n}}{p_s^{\rm Ent}(\omega,n)-2^{-n}} \ . 
\end{align}
We find the minimum $\nu_{\rm crit}^{\rm Ent}$ at low energies, i.e.~when $\omega \rightarrow 0$. In that limit, one sees that $\nu_{\rm crit}^{\rm Ent}\rightarrow \sqrt{\frac{n}{2^n-1}}$ decays exponentially towards zero as the number of parties $n$ grows, i.e.~the entanglement advantage has an exponential robustness to white noise.

\section{Certification of entanglement structure}
\label{sec:certification}

We now arrive at the main application of the distributed state discrimination task. Our framework enables the certification of entanglement in ensembles of energy-constrained quantum states using only a single separable measurement setting. The key insight is that entangled state preparations can generate correlations that are inaccessible to any fully separable ensemble under the same energetic restriction. Thus, observing a performance in the discrimination game that exceeds the separable bound constitutes a direct and operational witness of entanglement. 

In the multipartite setting, entanglement can manifest in a variety of forms. Many-body systems can exhibit different entanglement structures, ranging from partial entanglement within subsets of parties to genuine multipartite entanglement involving all parties. A commonly used entanglement quantificator in multipartite scenarios is $k$-separability \cite{Horodecki2009}. A quantum state distributed among $n$ parties is said to be $k$-separable if it can be written as the tensor product of $k$ pure states from $k$ disjoint subsets. The notion of $k$-separability, however, not always provides a full picture of the underlying entanglement structure. For example, the two quantum states $\ket{\sigma}=\ket{\phi}_{ABC}\otimes\ket{\psi}_{D}$ and $\ket{\sigma'}=\ket{\phi'}_{AB}\otimes\ket{\psi'}_{CD}$ are both $2$-separable, but the amount of entanglement to generate $\ket{\sigma}$ is larger, since it contains three-body entanglement. To distinguish between both cases, one can use the notion of producibility. A quantum state is said to be $l$-producible if all product states are at most of $l$ particles \cite{Guhne2005}. In this sense, the state $\ket{\sigma}$ is $2$-separable and $3$-producible, while $\ket{\sigma'}$ is $2$-separable and $2$-producible. There are, however, entanglement structures that still cannot be distinguished by both separability and producibility combined. The minimal example is found in a quantum sate composed by $n=7$ particles. Namely, the states $\ket{\tau}=\ket{\varphi}_{ABC}\otimes\ket{\xi}_{DEF}\otimes\ket{\chi}_{G}$ and $\ket{\tau'}=\ket{\varphi'}_{ABC}\otimes\ket{\xi'}_{DE}\otimes\ket{\chi'}_{FG}$ are both $3$-separable and $3$-producible. Their entanglement structure, however, is visibly not equivalent, since $\ket{\tau}$ requires two three-body entanglement while $\ket{\tau'}$ requires only one.

\begin{figure*}
\centering
\includegraphics[width=\textwidth]{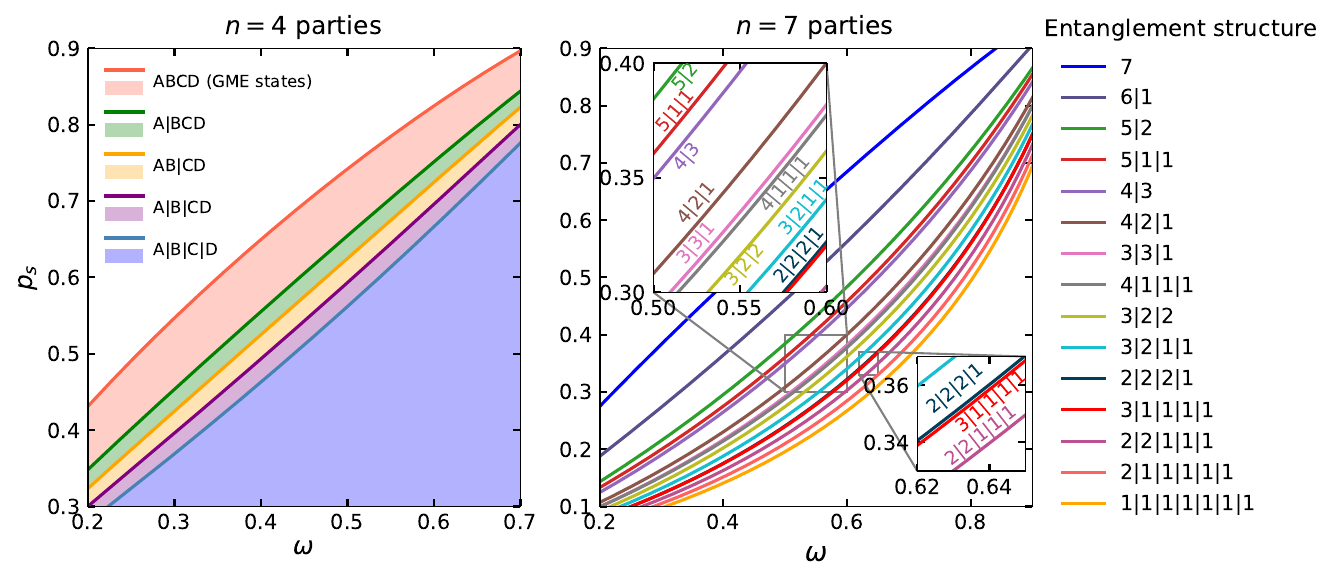}
\caption{\textit{Certification of the entanglement structure.} An ensemble of quantum states with energy $\omega$ is distributed among $n=4$ and $n=7$ parties. The success probability $p_s$ in the distributed state discrimination game depends on the structure of the entanglement. The entanglement structures on the right-hand side are ordered from most to least powerful.}
\label{fig:GME_plot}
\vspace{-0.5cm}
\end{figure*}

Our findings so far indicate that different forms of entanglement structures, independently of standard quantifiers such as their separability or producibility, lead to distinct levels of performance in the distributed state discrimination task. Specifically, ensembles with a stronger entanglement structure are able to achieve strictly larger success probabilities than those composed of states with shallower entanglement. This observation suggests that the task not only certifies the presence of entanglement, but can also quantify its structure in an operational way beyond standard approaches.

Following this intuition, we now investigate whether the distributed discrimination framework can also certify intermediate forms of multipartite entanglement, i.e., situations in which the ensemble is separable across certain bipartitions but entangled within others. To this end, we derive our final and most general result, which characterizes the maximum attainable success probability as a function of the underlying entanglement structure. \\

\begin{result} \label{result3}
Consider an ensemble of $2^n$ equiprobable quantum states $\rho_{\vec{x}}$ such that $\Tr\left[H \rho_{\vec{x}}\right]\leq \omega$. Let $\rho_{\vec{x}}$ be separable across $k$ groups of $r_j$ parties (for $j=0,\ldots,k-1$). The maximum probability that each party successfully discriminates their respective symbol $x_i\in\{0,1\}$ is
\begin{align} \label{eq:ps_gme}
p_s = \max_{\{\tilde{\omega}_j\}}\prod_{j=0}^{k-1} p_s^{\rm Ent}(\tilde{\omega}_j,r_j) \ ,
\end{align}
for $\prod_{j=0}^{k-1} \left(1-\tilde{\omega}_j\right) = 1-\omega$ and $\sum_{j=0}^{k-1}r_j=n$.
\end{result}
\begin{proof}
Let us express the global state as $\rho_{\vec{x}}=\bigotimes^{k-1}_{j=0}\sigma_{\vec{x}_j}^j$, where $\sigma_{\vec{x}_j}^j$ denotes the entangled quantum state shared among the $r_j$ parties forming the $j^{\rm th}$ group, and $\vec{x}_j$ collects all input symbols corresponding to those parties. Under this structure, the distributed discrimination task naturally decomposes into $k$ independent sub-tasks, each corresponding to the discrimination of the states $\sigma_{\vec{x}_j}^j$ within the respective group. Invoking Result~\ref{result1}, we know that the maximal success probability for discriminating the ensemble $\{\sigma_{\vec{x}_j}^j\}$ subject to an energy constraint $\tilde{\omega}_j$ is given by $p_s^{\rm Ent}(\tilde{\omega}_j,r_j)$. Since the global measurement strategy factorizes across the different groups, the overall success probability is simply the product of the optimal values for each subsystem, yielding $p_s=\prod_{j=0}^{k-1} p_s^{\rm Ent}(\tilde{\omega}_j,r_j)$. The individual energy parameters $\tilde{\omega}_j$ are not independent, as they must satisfy the global energy constraint $\prod_{j=0}^{k-1} \left(1-\tilde{\omega}_j\right) = 1-\omega$ which enforces conservation of the total energy across all groups. The remaining step consists in maximizing $p_s$ over all admissible sets $\{\tilde{\omega}_j\}$ consistent with the above constraint. Combining these elements, we recover \eqref{eq:ps_gme} as the expression for the maximal success probability achievable under partial separability.
\end{proof}

Although an analytic expression for the optimal value of \eqref{eq:ps_gme} is algebraically cumbersome, the optimization can be efficiently carried out numerically. Moreover, the concavity of the relevant functions ensures that the obtained extremum corresponds to the global maximum. There are, however, special cases in which the optimal $\{\tilde{\omega}_j\}$ become analytically simple. We show these cases in the following corollary.

\begin{corollary} \label{corollary1}
Let $\rho_{\vec{x}}$ be entangled across groups of the exact same number of parties, i.e.~$r_i=r_j:=r$, $\forall i,j$. The maximum probability for each party to successfully discriminate their corresponding symbol $x_i\in\{0,1\}$ is that one from \eqref{eq:ps_gme} with $\tilde{\omega}_j=1-\left(1-\omega\right)^{r/k}$, $\forall j$.
\end{corollary}
\begin{proof}
When all entangled groups contain the exact same number of parties, Lagrangian optimisation leads to a relation that is analogous to \eqref{eq:lagrangian_eqs}. It thus follows that all energy parameters $\tilde{\omega}_j$ are identical at the optimal point, resulting in $\tilde{\omega}_j=1-\left(1-\omega\right)^{r/k}$, $\forall j$.
\end{proof}

To illustrate the application of our method, let us consider a concrete example involving $n=4$ distant parties---Alice, Bob, Charlie, and Debbie---who share a family of quantum states $\{\rho_{\vec{x}}\}$, labelled by the classical bit-string $\vec{x} = (x_0, x_1, x_2, x_3)$. In the distributed state discrimination task, each party aims to infer their respective input bit: Alice guesses $x_0$, Bob guesses $x_1$, Charlie guesses $x_2$, and Debbie guesses $x_3$. Suppose that the ensemble $\rho_{\vec{x}}$ is bi-separable across the partition $A|BCD$. The optimal global set of states can then be written as $\rho_{\vec{x}} = \sigma_{x_0}^{A} \otimes \sigma_{x_1 x_2 x_3}^{BCD}$, where $\sigma_{x_0}^{A}$ acts locally on Alice’s system, while $\sigma_{x_1 x_2 x_3}^{BCD}$ represents an entangled state shared among Bob, Charlie, and Debbie. In this case, the overall success probability of the distributed discrimination task factorises as $p_s^{A|BCD} = p_s^{A} p_s^{BCD}$. According to Result~\ref{result3}, the optimal success probability for each group is given by
\begin{align}
p_s^{A} = p_s^{\rm Ent}(\tilde{\omega}_{A}, 1), \qquad 
p_s^{BCD} = p_s^{\rm Ent}(\tilde{\omega}_{BCD}, 3),
\end{align}
where $\tilde{\omega}_{A}$ and $\tilde{\omega}_{BCD}$ denote the portions of the total available energy $\omega$ assigned to each subsystem. Energy conservation imposes the constraint $(1 - \tilde{\omega}_{A})(1 - \tilde{\omega}_{BCD}) = 1 - \omega$. Maximising $p_s^{A|BCD} = p_s^{A} p_s^{BCD}$ under this constraint yields the optimal success probability for the given bipartition. Although no closed-form expression is available for the optimal values of $\tilde{\omega}_{A}$ and $\tilde{\omega}_{BCD}$, the maximisation can be efficiently performed numerically for any chosen total energy $\omega$. The concavity of the objective function guarantees that this numerical optimisation reaches the global maximum up to numerical precision.

We apply the same reasoning to the case in which the state is bi-separable across the finer partition $A|B|CD$. Here, the global success probability becomes a product of three terms, corresponding to the $A$-, $B$-, and $CD$-party groups, respectively. The remaining relevant configurations---namely $AB|CD$, the fully separable case ($A|B|C|D$), and the genuinely multipartite entangled case ($ABCD$)---all contain genuinely entangled partitions with the exact same number of parties, and can be derived directly from Result~\ref{result3} together with Corollary~\ref{corollary1}.  

The corresponding maximal success probabilities for each entanglement structure are displayed in \figref{fig:GME_plot} (left-hand side). Each curve defines a distinct entanglement structure boundary, specifying the maximal discrimination performance attainable by ensembles compatible with a given separability structure. The results reveal a clear hierarchy in the attainable success probabilities, reflecting the increasing strength of correlations permitted by more powerful entanglement structures. Explicitly, denoting by $\mathcal{P}(\cdot)$ the set of achievable success probabilities for each class of multipartite separability, we obtain the strict inclusions
\begin{align}
\mathcal{P}(A|B|C|D) &\subset \mathcal{P}(A|B|CD) \subset \mathcal{P}(AB|CD) \nonumber \\
&\subset \mathcal{P}(A|BCD) \subset \mathcal{P}(ABCD) \, .
\end{align}
As we allow stronger forms of entanglement, the system can produce a wider range of correlations between measurement outcomes. This expansion of possible correlations directly translates into a higher achievable success probability in the discrimination task. Operationally, this means that observing a success probability above a given bound immediately certifies that the underlying ensemble cannot belong to the corresponding separability class. In particular, surpassing the $A|BCD$ threshold (depicted as the green curve in \figref{fig:GME_plot}) constitutes unambiguous evidence of genuine multipartite entanglement, since no bi-separable decomposition, regardless of the chosen bipartition, can reproduce such correlations. 

We finally consider the case of ensembles of states distributed among $n=7$ parties. In \figref{fig:GME_plot} (right-hand side) we illustrate the maximum success probabilities for all possible inequivalent entanglement structures. Here, we denote by, e.g., $4|2|1$ a $3$-separable entanglement structure containing groups of $4$, $2$ and $1$ genuinely entangled particles in each disjoint group, respectively. Each entanglement structure yields a distinct maximum success probability, which is greater the stronger the entanglement structure in a hierarchical manner. For example, a GME state with entanglement structure $7$ yields the greatest maximum success probability. The second most powerful entanglement structure is $6|1$ above which, the ensemble of states is certified to be GME. The complete list of entanglement structures ordered from most to least powerful can be found on the right-hand side of \figref{fig:GME_plot}. Interestingly, one can see how some of the structures distinguished by this method can be missed by looking solely at both their separability and producibility. For example, the entanglement structures $3|3|1$ and $3|2|2$ represent states that are both $3$-separable and $3$-producible. However, the distributed state discrimination game detects that the structure $3|3|1$ is able to produce more powerful correlations, and thus generate more distinguishable ensembles, than the structure $3|2|2$. The distributed state discrimination task thus provides a fully operational and experimentally accessible method to certify not only the presence of entanglement, but also its underlying structure, directly from observable success probabilities.

\section{Conclusion}

We have introduced a semi-device-independent framework for certifying the underlying entanglement structure of energy-restricted quantum systems, grounded on a distributed state discrimination game played among distant and uncharacterised parties. In this game, each participant attempts to decode a party-specific bit from a shared ensemble of quantum states. We demonstrated that entangled preparations outperform all separable strategies under identical energy constraints, and that this advantage increases systematically with the entanglement depth of the ensemble. Consequently, the observed success probability serves as an operational and quantitative signature of the structure of the multipartite entanglement beyond standard structure quantifiers such as separability of producibility. A key strength of the approach is its simplicity: the protocol relies solely on separable measurements, each with a single setting, and remains robust under realistic experimental imperfections such as white noise. Moreover, the protocol is semi-device-independent, relying on a physically testable assumption such as the energy of the prepared ensemble. These treats contrast with entanglement-structure certification approaches which rely on the hardly testable assumption of a limited Hilbert space dimension, or device-independent protocols that require Bell-like violations and multiple measurement settings. All together makes the proposed method directly amenable to implementation in existing quantum optical and solid-state platforms. 

Looking ahead, several research directions emerge. A particularly promising avenue involves experimental realisations with photonic systems, for example using multimode squeezed light, where energy constraints arise naturally and the required measurements are already experimentally accessible. Our framework also provides a new tool for probing non-classical correlations in Gaussian states, which represent key resources in continuous-variable quantum information processing. Extending the analysis to scenarios with a larger number of inputs and outcomes could enable a richer characterization of multipartite correlations, including not only entanglement structure but also its dimensionality \cite{cobucci2024}. Our method could then directly compete with alternative entanglement structure quantificators based on the Schmidt number \cite{eisert2001,Shahandeh2014}. Beyond its foundational implications, the distributed state discrimination game may find applications in quantum communication and quantum key distribution, where energy constraints are both natural and operationally relevant. In these contexts, the ability to certify entanglement depth using only minimal device assumptions offers a practical path toward more secure and resource-efficient quantum communication.

\begin{acknowledgements} 
The author thanks Nicola D'Alessandro and Armin Tavakoli for discussions and constructive feedback on the manuscript. This work is supported by the Wenner-Gren Foundations.
\end{acknowledgements}

\bibliography{d_qsd}

\end{document}